\documentclass[aps,prx,twocolumn,superscriptaddress,showpacs,floatfix]{revtex4-1}
\usepackage{amsmath,amssymb,amsfonts,float,graphics,epsfig,epstopdf,color,verbatim,tabularx,bm}
\usepackage[colorlinks,linkcolor=blue,citecolor=blue,urlcolor=blue]{hyperref}
\usepackage{wasysym}
\usepackage{color}
\newcommand{\ii}{\mathrm{i}}
\newcommand{\dd}{\mathrm{d}}
\newcommand{\scL}{\mathcal{L}}
\newcommand{\vect}[1]{{\bm{#1}}}

\renewcommand{\Im}{\mathrm{Im}}
\newcommand{\eqnref}[1]{Eq.\,\eqref{#1}}
\newcommand{\figref}[1]{Fig.\,\ref{#1}}
\newcommand{\tabref}[1]{Tab.\,\ref{#1}}

\newcommand{\beq}{\begin{equation}}
\newcommand{\eeq}{\end{equation}}
\newcommand{\beqn}{\begin{eqnarray}}
\newcommand{\eeqn}{\end{eqnarray}}
\graphicspath{{Figures/}}

\begin{document}

\title{Dynamical Signature of Fractionalization at a Deconfined Quantum Critical Point}

\author{Nvsen Ma}
\affiliation{Beijing National Laboratory for Condensed Matter Physics and Institute of Physics, Chinese Academy of Sciences, Beijing 100190, China}

\author{Guang-Yu Sun}
\affiliation{Beijing National Laboratory for Condensed Matter Physics and Institute of Physics, Chinese Academy of Sciences, Beijing 100190, China}
\affiliation{School of Physical Sciences, University of Chinese Academy of Sciences, Beijing 100190, China}

\author{Yi-Zhuang You}
\affiliation{Department of Physics, Harvard University, Cambridge, MA 02138, USA}
\affiliation{Department of Physics, University of California, San Diego, California 92093, USA}

\author{Cenke Xu}
\affiliation{Department of physics, University of California, Santa Barbara, CA 93106, USA}

\author{Ashvin Vishwanath}
\affiliation{Department of Physics, Harvard University, Cambridge, MA 02138, USA}

\author{Anders W. Sandvik}
\affiliation{Beijing National Laboratory for Condensed Matter Physics and Institute of Physics, Chinese Academy of Sciences, Beijing 100190, China}
\affiliation{Department of Physics, Boston University, 590 Commonwealth Avenue, Boston, Massachusetts 02215, USA}

\author{Zi Yang Meng}
\affiliation{Beijing National Laboratory for Condensed Matter Physics and Institute of Physics, Chinese Academy of Sciences, Beijing 100190, China}
\affiliation{CAS Center of Excellence in Topological Quantum Computation and School of Physical Sciences, University of Chinese Academy of Sciences, Beijing 100190, China}
\affiliation{Songshan Lake Materials Laboratory, Dongguan, Guangdong 523808, China}

\date{\today}

\begin{abstract}
Deconfined quantum critical points govern continuous quantum phase transitions at which fractionalized (deconfined) degrees of freedom emerge. Here we study dynamical signatures of the fractionalized excitations in a quantum magnet (the easy-plane J-Q model) that realize a deconfined quantum critical point with emergent O(4) symmetry. By means of large-scale quantum Monte Carlo simulations and stochastic analytic continuation of imaginary-time correlation functions, we obtain the dynamic spin structure factors in the $S^{x}$ and $S^{z}$ channels. In both channels, we observe broad continua that originate from the deconfined excitations. We further identify several distinct spectral features of the deconfined quantum critical point, including the lower edge of the continuum 
and its form factor on moving through the Brillouin Zone. We  provide field-theoretical and lattice model calculations that explain the overall shapes of the computed spectra, which highlight the importance of interactions and gauge fluctuations to explaining the spectral-weight distribution. We make further comparisons with the conventional Landau  O(2) transition in a different quantum magnet, at which no signatures of fractionalization are observed. The distinctive spectral signatures of the deconfined quantum critical point suggest the feasibility of its experimental detection in neutron scattering and nuclear magnetic resonance experiments.
\end{abstract}

\maketitle

\section{Introduction}

\label{sec:introduction}

The deconfined quantum critical point (DQCP), which separates the N\'eel antiferromagnetic (AFM) and spontaneously dimerized valence bond 
solid (VBS) phases in (2+1)D quantum magnets, was proposed as an example of continuous quantum phase transition outside the conventional 
Landau-Ginzburg-Wilson (LGW) paradigm \cite{deconfine1,deconfine2}. The AFM and VBS order parameters both vanish continuously and simultaneously 
at the DQCP. This scenario is generically not expected within the standard LGW description, where such a case should be realizable only by fine 
tuning two separate transitions to coincide at special multi-critical points. Multiple field theory descriptions
\cite{ashvinlesik,deconfine1,deconfine2,hermele2005,ranwen,senthilfisher,groverashvin,lulee,xudual,karchtong,seiberg2,mengxu,SO5,maxryan17,youSMG2} 
have been proposed for the DQCP which are believed to be equivalent (or dual) to each other at low energy, including the non-compact CP$^1$ (NCCP$^1$) 
theory \cite{deconfine1,deconfine2} and some versions of the quantum electrodynamics (QED) and quantum chromodynamics (QCD) theories \cite{SO5,2018arXiv180707574X}. 
In contrast to the LGW description which formulates the critical theory in terms of the order parameters directly, these gauge theory descriptions 
of the DQCP are formulated in terms of deconfined degrees of freedom (fractionalized particles and emergent gauge fields). The order parameters 
on either side of the DQCP can be expressed as different compositions of the fractionalized particles or gauge fluctuations within the same 
theoretical framework. This mechanism captures the intertwinement of the AFM and VBS orders and provides a natural route beyond the LGW paradigm
to a non-fine-tuned quantum critical point between the two distinct symmetry-breaking phases. 

With the increasing understanding of the nature of the DQCP ground state phase transition, the time is now ripe to address direct connections to 
experiments, where the most detailed signatures of deconfinement can be expected in dynamical properties. Based on the physical picture of deconfinement
of the experimentally accessible spin excitation into two spinons at the DQCP, a broad continuum is expected in the spectral function. This is in 
sharp contrast to an LGW transition of the AFM state into a nondegenerate (trivial) quantum paramagnet, where spinwaves (magnons) picture remain approximately valid at the critical point (as a very sharp edge of the critical continuum, albeit the magnon quasiparticle weight is highly damped to zero) \cite{Chubukov1994}. The aim of this paper is to present a comprehensive numerical 
study of the signature of magnon fractionalization in the dynamic spin structure factor $S(\vect{q},\omega)$ of a (2+1)D square-lattice spin 
model hosting a DQCP, accompanied with a detailed field theory analysis of every low-energy continuum that appears in the spectrum.

Following the DQCP  proposal, intensive theoretical and numerical efforts have been invested in the possibility of unambiguously observing 
such critical points in lattice models. In the traditional frustrated quantum spin models that exhibit VBS phases, sign problems in quantum 
Monte Carlo (QMC) simulations and other technical difficulties in methods such as the density matrix renormalization group and tensor product states 
prohibit studies of the large system sizes needed in order to reliably characterize critical points. However, for generic and universal 
properties, other ``designer hamiltonians'' \cite{Kaul2013a} can be constructed that do not suffer from QMC sign problems but still host the 
desired phases. Many such studies have pointed to the existence of the DQCP in both two-dimensional (2D) quantum magnets  
\cite{Sandvik2002,Sandvik2007,Lou2009,Sandvik2010,Pujar2013,Block2013,Shao2016,Jonathan2017,YQQin2017,XFZhang2017} and related (through the 
path integral) three-dimensional (3D) classical models~\cite{Nahum2011,Nahum2015a,Nahum2015b,Sreejith18}. In these studies it has been observed, e.g., 
that the order parameters have unusually large anomalous dimensions~\cite{Lou2009,Block2013,Nahum2015b,Shao2016,YQQin2017,XFZhang2017}, which 
is an important deviation from the common 3D Wilson-Fisher fixed point. More concrete evidence of deconfinement has been found by directly probing 
the length scale associated with the fractionalization process \cite{Tang2013,Shao2016} and from thermodynamics \cite{Sandvik2011}. However, 
the experimentally most direct signatures of a DQCP, the dynamic spin structure factor $S(\vect{q},\omega)$, have so far not been calculated in the 
case of electronic spins (while there are already some intriguing results for an SU(3) symmetric model~\cite{Assaad2016}). Historically, in 
quasi-1D systems, the experimentally observed spinon continuum, which agrees with calculations for the spin-$1/2$ Heisenberg chain, was crucial 
in establishing spinon deconfinement. Indications of fractionalized magnetic excitations in 2D quantum spin liquids have also been similarly 
observed \cite{HanTH12,FuM15,FengZL17,WenXG17,WeiYuan2017,ZLFeng2017dopeZn,GYSun2018}. Given that $S(\vect{q},\omega)$ is detectable by multiple experimental 
techniques, including  inelastic neutron scattering (INS), resonant inelastic X-ray scattering (RIXS) and nuclear magnetic resonance (NMR), 
identifying the distinct signatures of fractionalization in $S(\vect{q},\omega)$ at the DQCP will provide a useful guide to experimental searches 
for DQCPs in magnetic materials. Since the qualitative features of $S(\vect{q},\omega)$ remain the same in the entire critical ``fan'' extending from the critical point to finite temperature, the dynamical signatures proposed in our study should be robustly observed even if the experimental parameter is slightly off the critical point.
Moreover, due to a recently investigated duality relation between the DQCP and a certain bosonic topological 
transitions in fermion systems \cite{SO5,YQQin2017,XFZhang2017}, similar dynamical signature of fractionalization is also expected in 
interaction-driven topological phase transitions. Therefore our work also can impact the ongoing efforts in finding experimentally accessible 
signatures of topological phase transitions in strongly correlated electron systems.

\begin{figure}[t]
\includegraphics[width=\columnwidth]{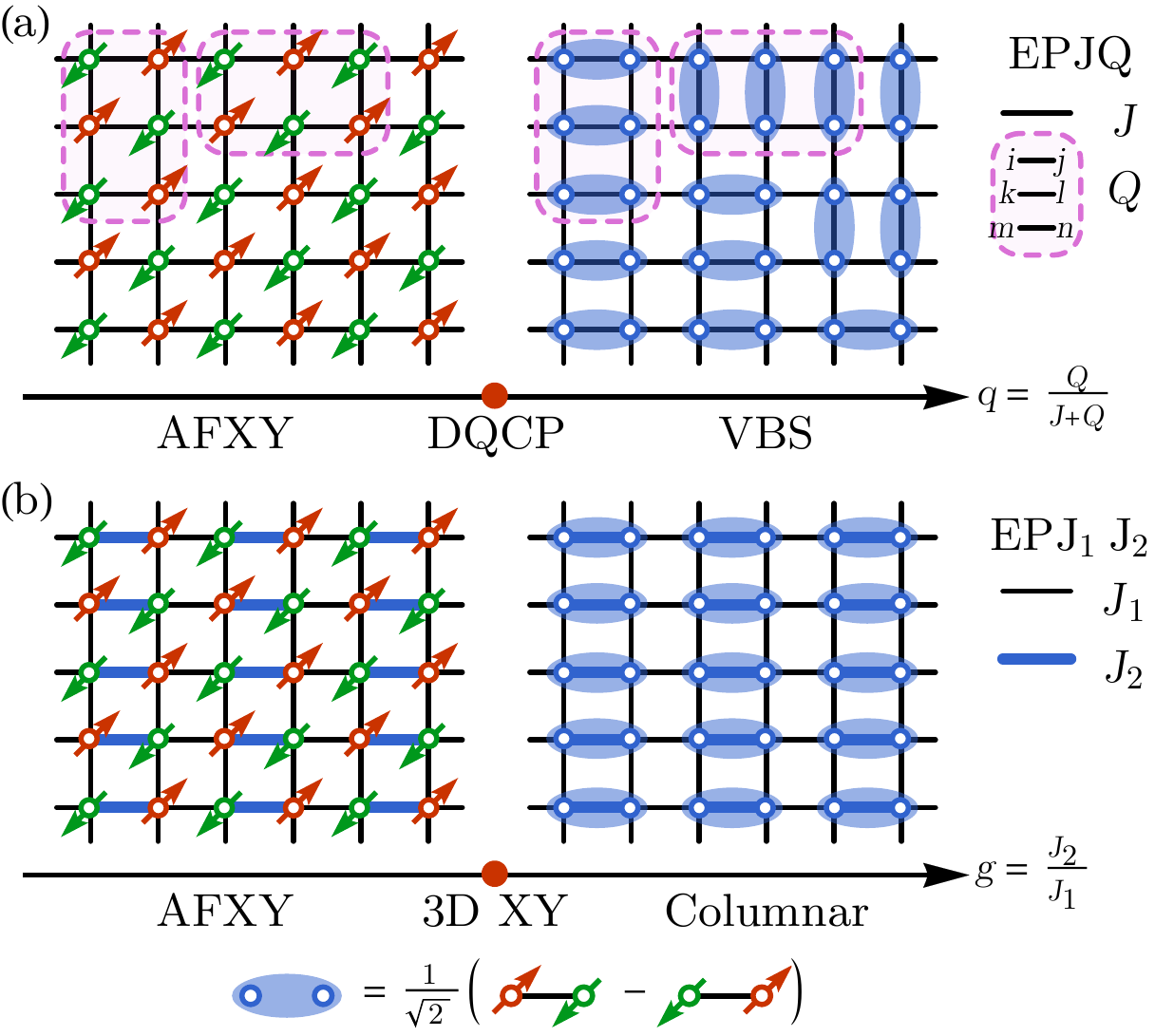}
\caption{The two lattice models considered in this work and their schematic phase diagrams. (a) The EPJQ model with two-spin ($J$) and six-spin 
($Q$) couplings preserve all symmetries of the square lattice. We define the tuning parameter chosen as $q=Q/(J+Q)$. The antiferromagnetic XY 
(AFXY) phase is separated by the DQCP at $q=q_c$ from the columnar VBS phase, which spontaneously breaks lattice symmetries but which has significant 
fluctuations of the four-fold degenerate dimer pattern close to $q_c$, as indicated. (b) The EPJ$_1$J$_2$ model, with the tuning parameter $g=J_2/J_1$. 
The $J_2$ term explicitly pins a columnar dimer pattern and drives the AFXY phase to the spin-disordered trivial (non-degenerate) columnar singlet 
phase (without spontaneous lattice symmetry breaking) through the 3DXY transition at $g=g_c$.} \label{fig:model}
\end{figure}

In this work, we will investigate a U(1) version of the DQCP on the square lattice, with the easy-plane $J$-$Q$ (EPJQ) model defined 
by the Hamiltonian
\begin{equation}\label{eq:EPJQ}
H_\text{JQ}=-J\sum_{\langle ij\rangle}(P_{ij}+\Delta S_i^zS_j^z)-Q\hskip-2mm\sum_{\langle ijklmn \rangle}\hskip-2mm
P_{ij}P_{kl}P_{mn},
\end{equation}
were $\vect{S}_i$ denotes the spin-1/2 operator on each site $i$ and $P_{ij}=\tfrac{1}{4}-\boldsymbol{S}_i\cdot\boldsymbol{S}_j$ is the 
singlet-projection operator on the link $ij$ (between nearest-neighbor sites). The two- and six-spin terms are both illustrated in 
\figref{fig:model}(a). For $\Delta=0$ this is the previously studied SU(2)$_\text{spin}$ $J$-$Q_3$ model~\cite{Lou2009,Sen2010,Sandvik2010}, which 
is an extension of the original $J$-$Q$ model (or $J$-$Q_2$ model) \cite{Sandvik2007}, with two instead of three singlet projectors in the $Q$ terms. 
With three singlet 
projectors we can go further into the VBS state while still keeping $J>0$ in sign-free QMC simulations. The term $\Delta S_i^zS_j^z$ with 
$\Delta \in (0,1]$ introduces the easy-plane anisotropy that breaks the $\text{SU}(2)_\text{spin}$ symmetry down to $\text{U}(1)_\text{spin}$ 
explicitly. It has been shown \cite{YQQin2017} that when $\Delta=1/2$ (which is the value we will use here), 
the EPJQ model exhibits a direct and continuous quantum phase transition 
between the AFXY and VBS phases, as illustrated in \figref{fig:model}(a), realizing the easy-plane DQCP (while for larger anisotropy, such as 
$\Delta=1$, the transition becomes first-order). The XY order parameter has a U(1)$_\text{spin}$ rotational symmetry and the VBS order parameter 
exhibits an emergent U(1)$_\text{VBS}$ symmetry as the DQCP is approached, and, as argued based on dualities \cite{SO5}, the two U(1) symmetries 
combine to form an emergent higher O(4) symmetry exactly at the DQCP.

To make a comparison with the EPJQ model, we will also study an easy-plane $J_1$-$J_2$ (EPJ$_1$J$_2$) model,
\begin{equation}\label{eq:EPJ1J2}
H_\text{J$_1$J$_2$}=J_1\sum_{\langle i,j \rangle'}D_{ij} +J_2\sum_{\langle i,j \rangle''}D_{ij},
\end{equation}
where $D_{ij}=S^{x}_{i}S^{x}_j + S^{y}_{i}S^{y}_{j}+\Delta  S^{z}_{i}S^{z}_j$. The $J_1$ bonds $\langle i,j \rangle'$ and the $J_2$
bonds $\langle i,j \rangle''$ correspond to the thin black and the thick blue bonds in \figref{fig:model}(b) respectively. Since the Hamiltonian
explicitly breaks the lattice symmetry, with the $J_2$ terms pinning a columnar pattern of bonds with higher singlet density, the large $J_2$ 
phase will simply be a trivial, non-degenerate quantum paramagnet. The transition out of the AFXY phase is then the conventional O(2) Wilson-Fisher 
transition in the 3D XY universality class, as illustrated in  \figref{fig:model}(b), which we will contrast with the DQCP. We here take $\Delta=1/2$ 
for the anisotropy parameter.

By means of stochastic analytic continuation (SAC) of imaginary-time correlation functions calculated using large-scale QMC simulations, 
we extract the dynamic spin structure factor $S(\vect{q},\omega)$ over a wide range of momentum ($\vect{q}$) and energy ($\omega$) transfers in 
both the EPJQ and EPJ$_1$J$_2$ models. The calculations are performed in all phases of the models as well as at the critical points for both the $S^x$ and $S^z$ spin channels.
Our result confirms the broad spinon continua in both $S^x$ and $S^z$ channels at the DQCP, as expected from the spinon deconfinement. We also identify the following distinctive spectral features: (i) the lower edge of the spinon continua well fitted by a simple function $(\sin^2 (q_x)+\sin^2 (q_y))^{1/2}$ which reveals the spinon dispersion relation, (ii) six gapless continua at $S^z(\pi,\pi)$,  $S^x(0,0)$, $S^x(\pi,0)$, $S^x(0,\pi)$, $S^z(\pi,0)$, $S^z(0,\pi)$ unique to the DQCP, (iii) the characteristic spatial anisotropy of the $(\pi,0)$ and $(0,\pi)$ continua, (iv) the remarkable similarity between $S^x$ and $S^z$ spectrums despite of the strong easy-plane anisotropy. We will also show that these specific signatures of deconfined spinons can be observed in the EPJQ model over a wide range of model parameters even away from the exact critical point, demonstrating that the dynamic signatures we identified are robust phenomena that should be testable in future experiments.

The rest of the paper is organized as follows: In Sec.~\ref{sec:theories}, the theoretical background of the DQCP phenomenon and the consequently 
expected low energy spectral features are laid out. In Sec.~\ref{sec:spectra}, we discuss in detail the dynamic spin structure factors in the EPJQ 
and EPJ$_1$J$_2$ models, as they are driven through their phase transitions. This comparison reveals the distinct spectral 
features of the DQCP. In Sec.~\ref{sec:meanfield}, we provide a theoretical calculation of the dynamic spin structure factor at the DQCP which
nicely matches the numerical observations. Sec.~\ref{sec:discussion} summarizes the significance of our findings and their relevance to bridging 
the DQCP to experimentally accessible information, and points out future directions. A detailed finite-size scaling analysis of the not previously 
studied quantum phase transition of the EPJ$_1$J$_2$ model and additional discussion of the spin spectra of this model are given in Appendix~\ref{sec:EPJ1J2QCP} and ~\ref{sec:EPJ1J2completespectra}, respectively. 

\section{Theoretical Expectations for Low-Energy Spectral Features} \label{sec:theories}

\begin{figure}[tp]
\begin{center}
\includegraphics[width=0.9\columnwidth]{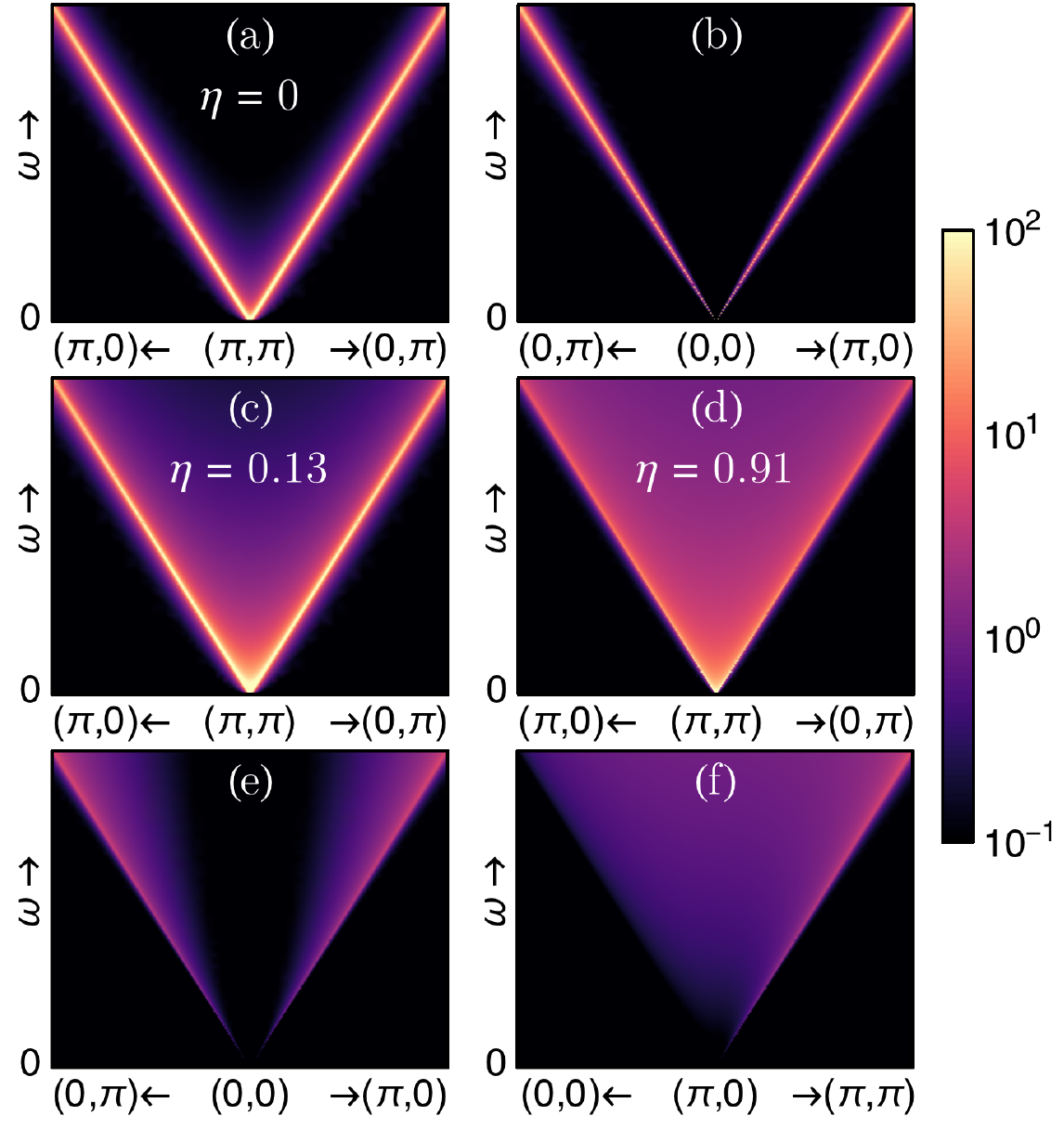}
\caption{Expected low-energy features of the dynamic spin structure factor $S(\vect{q},\omega)\equiv\Im \chi(\vect{q},\omega+\ii 0_+)$ based 
on the theoretical dynamic spin susceptibility $\chi(\vect{q},\omega)$ listed in \tabref{tab:features}.} \label{fig:features}
\end{center}
\end{figure}

Before presenting our numerical result, we would like to first provide a theoretical overview of the expected spectral features at low energy, 
as summarized in \tabref{tab:features} and \figref{fig:features}. Let us define the N\'eel AFM $\vect{n}=(n_x,n_y,n_z)$ and VBS $\vect{v}=(v_x,v_y)$ order 
parameters as
\begin{equation}
\vect{n}=(-)^i \vect{S}_i, \quad v_a=(-)^a \vect{S}_{i}\cdot\vect{S}_{i+\hat{a}}.
\end{equation}
With the easy-plane anisotropy, the XY order parameter in the AFXY phase is just the planar component $(n_x,n_y)$ of the N\'eel order parameter $\vect{n}$.

Deep in the AFXY phase of both the EPJQ and EPJ$_1$J$_2$ models, the low-energy fluctuations are described by the XY 
model, $\mathcal{L}[\theta]=(\rho_s/2)(\partial \theta)^2$, where $\rho_s$ is the spin stiffness and $\theta$ is the spin-wave Goldstone 
mode, such that the XY order parameter can be written as $n_x+\ii n_y\sim e^{\ii \theta}$. The XY spin correlation function $S^{x}$ near momentum 
$(\pi,\pi)$ is expected to follow
\begin{equation}
\langle e^{-\ii\theta} e^{\ii
\theta}\rangle\sim \rho_s \delta(q) +
(\vect{q}^2-\omega^2)^{-1}+\cdots,
\end{equation}
where $q = (\omega, \vect{q})$. The imaginary part of this correlation function (with $\omega\to\omega+\ii 0_+$) is shown in \figref{fig:features} (a), 
demonstrating the well-defined magnon mode with linear dispersion. On the other hand, the spin $S^z$ fluctuation is gapped at $(\pi,\pi)$ due to 
the easy-plane anisotropy, but becomes gapless at $(0,0)$. The excitation of $S^z$ corresponds to the spin density fluctuation 
$\partial_t\theta\sim n_x\partial_t n_y-n_y\partial_t n_x$, which can decay into two gapless magnon modes, each around $(\pi,\pi)$, so that
the total momentum is close to $(0,0)$. Therefore, we expect $S^{z}$ near $(0,0)$ to be of the form
\begin{equation}
\langle\partial_t\theta\partial_t\theta\rangle \sim
\frac{\omega^2}{\rho_s(\vect{q}^2-\omega^2)}. 
\end{equation}
The imaginary part of this correlation function (with $\omega\to\omega+\ii 0_+$) is shown in \figref{fig:features} (b). Here the spectral 
weight of the linearly dispersing mode is suppressed as $\omega\to 0$. As we will see, the low-energy spectral features of the dynamic spin structure factors $S^{x}(\vect{q},\omega)$ and $S^{z}(\vect{q},\omega)$ match our QMC-SAC results nicely [see (a) and (b) for both \figref{fig:EPJQ32} and \figref{fig:EPJ1J232}].

\begin{table}[t]
\caption{Analytical expressions for the low-energy dynamic spin susceptibility 
$\chi(\vect{q},\omega)\equiv\int_{-\infty}^{+\infty}(\dd \epsilon/\pi) (\omega-\epsilon)^{-1}S(\vect{q},\epsilon)$ close
to the gapless momentum points.  The physical meanings of all these low-energy modes are listed in the last column in terms 
of the AFM ($\vect{n}$) and VBS ($\vect{v}$) order parameters.}
\begin{center}
\begin{tabular}{ccccc}
& low energy $\chi(\vect{Q}+\vect{q},\omega)$ & channel & $\vect{Q}$ & mode\\
\hline
(a) & $(\vect{q}^2-\omega^2)^{-1}$ & AFXY $S^{x}$ & $(\pi,\pi)$ & $n_x$ \\
(b) & $\omega^2(\vect{q}^2-\omega^2)^{-1}$ & AFXY $S^{z}$ & $(0,0)$ & $n_x\partial_t n_y$ \\
(c) & $(\vect{q}^2-\omega^2)^{-1+\eta_{xy}/2}$ & DQCP $S^{x}$ & $(\pi,\pi)$ & $n_x$ \\
(d) & $(\vect{q}^2-\omega^2)^{-1+\eta_z/2}$ & DQCP $S^{z}$ & $(\pi,\pi)$ & $n_z$\\
(e) & $\vect{q}^2(\vect{q}^2-\omega^2)^{-1/2}$ & DQCP $S^{z}$ & $(0,0)$ & $n_x\partial_t n_y$\\
(f) & $(\omega^2-q_x^2)(\vect{q}^2-\omega^2)^{-1/2}$ & DQCP $S^{x}$ & $(\pi,0)$ & $n_x\partial_yv_y$
\end{tabular}
\end{center}
\label{tab:features}
\end{table}

At the DQCP, the low-energy dynamic spin susceptibility around $\vect{Q}=(\pi,\pi)$ is expected to be of the from
\begin{equation}
\begin{split}
\chi^{x}(\vect{Q}+\vect{q},\omega)&\sim(\vect{q}^2-\omega^2)^{-1+\eta_{xy}/2},\\
\chi^{z}(\vect{Q}+\vect{q},\omega)&\sim(\vect{q}^2-\omega^2)^{-1+\eta_{z}/2},
\end{split}
\end{equation}
with large values of the anomalous dimension $\eta_{xy}$ and $\eta_{z}$, characterizing the complete breakdown of a well-defined magnon at the 
critical point. The imaginary part of these correlation functions are shown in \figref{fig:features} (c,d) respectively, with $\eta_{xy}\approx0.13$ 
and $\eta_z\approx0.91$ taken from Ref.\,\cite{YQQin2017} for illustration purpose. Strictly speaking, any non-zero anomalous dimension $\eta$ 
would imply the breakdown of well-defined magnons, but compared to the small anomalous dimension $\eta\approx0.04$ \cite{zinnjustin98,campostrini2001} 
at the 3D O(2) Wilson-Fisher transition, we expect to observe a much more prominent continuum at the DQCP in the EQJQ model [as clearly seen in the
QMC-SAC results \figref{fig:EPJQ32}(b,e), to be discussed later]. 
This is in sharp contrast to the $S^{x}(\vect{q},\omega)$ spectrum at the 3DXY critical point in the EPJ$_1$J$_2$ model [as shown 
in \figref{fig:EPJ1J232}(b)], where there is essentially no continuum in the gapless $S^x$ channel and that in the gapped $S^z$ channel is much less
prominent (though there are also interesting features there that cannot be explained at the level of analysis discussed above).

Another important feature in the spectrum of the DQCP is the gapless excitations at momenta $(0,0)$ and $(\pi,0)$ [as well as $(0,\pi)$ by
symmetry] in both $S^{x}$ and $S^{z}$ channels, with much weaker spectral weight, as shown in \figref{fig:EPJQ32} (b,e). Theoretically, they 
correspond to the (generally non-conserved) SO(5) current fluctuations, where the SO(5) group rotates the N\'eel and VBS order parameters as 
a combined vector $\vect{N}=(N_1,N_2,N_3,N_4,N_5)=(\vect{n},\vect{v})$. The SO(5) current can be written in terms of the combined order parameter 
$\vect{N}$ as
\begin{equation}
j^\mu_{ab} = N_a\partial_\mu N_b-N_b\partial_\mu N_a,
\end{equation}
with $\mu=0,1,2$ and $a,b=1,\cdots,5$. By matching the momentum and the SO(5) symmetry quantum numbers, it is straightforward to identify the $S^{x}$ and $S^{z}$ fluctuations around 
$(0,0)$ to $j^0_{23}$ and $j^0_{12}$ respectively, and identify those around $(\pi,0)$ to $j^{2}_{15}$ and $j^2_{35}$ respectively. The 
emergent O(4) symmetry at the easy-plane DQCP corresponds to the subgroup of SO(5) that rotates $(N_1,N_2,N_4,N_5)$ only (keeping $N_3^2$ invariant), 
so the currents $j^0_{12}$ and $j^2_{15}$ are emergent conserved currents at low-energy. Their correlation functions can be calculated based 
on the $N_f=2$ QCD theory or the $N_f=4$ QED theory $\scL[\psi,a]=\bar{\psi}\gamma^\mu D_\mu\psi+\cdots$, where the order parameters are 
fractionalized as $N_a\sim\bar{\psi}\Gamma^a\psi$ and the current-current correlations are given by
\begin{equation}\label{eq:current1}
\begin{split}
\langle j^0_{12} j^0_{12}\rangle&\sim\langle\bar{\psi}\gamma^0\Gamma^{12}\psi\bar{\psi}\gamma^0\Gamma^{12}\psi\rangle\sim\frac{\vect{q}^2}{(\vect{q}^2-\omega^2)^{1/2}},\\
\langle j^2_{15} j^2_{15}\rangle&\sim\langle\bar{\psi}\gamma^2\Gamma^{15}\psi\bar{\psi}\gamma^2\Gamma^{15}\psi\rangle\sim\frac{\omega^2-q_x^2}{(\vect{q}^2-\omega^2)^{1/2}},
\end{split}
\end{equation}
with $\Gamma^{ab}=\frac{\ii}{2}[\Gamma^a,\Gamma^b]$ being the SO(5) generator that rotates $(N_a,N_b)$ components. These spectral functions of the
currents in the field theory correspond in the lattice model to the spin spectrum $S^{z}$ around $(0,0)$ and $S^{x}$ around $(\pi,0)$. The 
imaginary part of these correlation functions are show in \figref{fig:features} (e,f) respectively. 

The correlation function of the non-conserved 
currents $j^0_{23}$ and $j^2_{35}$ are expected to take a similar form with another anomalous dimension $\eta_j$,
\begin{equation}\label{eq:current2}
\begin{split}
\langle j^0_{23} j^0_{23}\rangle&\sim\frac{\vect{q}^2}{(\vect{q}^2-\omega^2)^{(1-\eta_j)/2}},\\
\langle j^2_{35} j^2_{35}\rangle&\sim\frac{\omega^2-q_x^2}{(\vect{q}^2-\omega^2)^{(1-\eta_j)/2}}.
\end{split}
\end{equation}
They correspond to $S^{x}(\vect{q},\omega)$ around $(0,0)$ and $S^{z}(\vect{q},\omega)$ around $(\pi,0)$. As we will discuss in more detail in Sec.~\ref{sec:spectra}
and Sec.~\ref{sec:meanfield}, all these expected spectral features are qualitatively observed in the QMC-SAC spectrum of the EPJQ model 
[see \figref{fig:EPJQ32} (b,e)], consistent with the QCD or QED description of the DQCP.

In the VBS phase, all excitations (in both $S^x$ and $S^z$ channels) are gapped. There is no low-energy feature in the spectrum that can be 
reliably predicted at the field theory level. With our QMC-SAC numerics we can easily go into the VBS, however, and we will present results 
along with the results in the XY phase and DQCP in the next section.

\section{Numerical Calculations of the Spin Spectra}
\label{sec:spectra}

We here present results for both the EPJQ and the EPJ$_1$J$_2$ models. The key quantity computed in our QMC simulations with the stochastic series expansion 
(SSE) method \cite{Sandvik2010b} is the  spin correlation function in the imaginary time domain (for $a=x,y,z$), 
\begin{equation}
\bar G^{a}(\vect{q},\tau)=\langle S^{a}_{-\vect{q}}(\tau)S^{a}_{\vect{q}}(0)\rangle,
\end{equation}
where $S^{a}_{\vect{q}}=\frac{1}{L}\sum_{i} e^{-\ii\vect{q}\cdot\vect{r}_i} S^{a}_{i}$ and the summation is over all sites of the $L\times L$ lattice. 
From the imaginary time data for a set of $\tau$ points, we reconstruct the corresponding real-frequency spectral function by performing a numerical
analytic continuation using the SAC method~\cite{Sandvik1998a,Beach2004,Syljuasen2008,Sandvik2015,Qin2017,Shao2017a,Shao2017b,Huang2017}. With
this method, we average over Monte Carlo importance-sampled spectral functions $B^{a}(\vect{q},\omega)$, from which the dynamic spin structure factor
is later obtained as $S^{a}(\vect{q},\omega) = \langle B^{a}(\vect{q},\omega)\rangle /(1+e^{-\beta\omega})$. The intermediate spectrum 
$B^{a}(\vect{q},\omega)$ has the advantage of being normalized to $\bar G^{a}(\vect{q},0)$ when integrating over positive frequencies only. In 
the sampling procedure we thus fix the normalization and use the relationship
\begin{equation}
G^{a}(\vect{q},\tau)=\int_0^{\infty}\frac{\dd \omega}{\pi}\frac{e^{-\tau\omega}+e^{-(\beta-\tau)\omega}}{1+e^{-\beta\omega}}B^{a}(\vect{q},\omega)
\end{equation}
to define the goodness-of-fit $\chi^2$ between this function and the SSE-computed result $\bar G^{a}(\vect{q},\tau)$ (including covariance among 
the SSE data for different $\tau$). The weight for a given spectrum is $\propto {\rm exp}({-\chi^2/2\theta})$, with $\theta$ a fictitious temperature 
chosen in an optimal way so as to give a statistically sound mean $\chi^2$ value, while still staying in the regime of significant fluctuations 
of the sampled spectra so that a smooth averaged spectral function is obtained. The most recent incarnation of the SAC method uses a parametrization 
with a large number of equal-amplitude $\delta$-function sampled at locations in a frequency continuum and collected in a histogram, as explained
in Refs.~\onlinecite{Qin2017,Shao2017a,Shao2017b,Huang2017,YRShu2017}. We refer to these works for technical details.

\begin{figure*}[htp!]
\includegraphics[width=\textwidth]{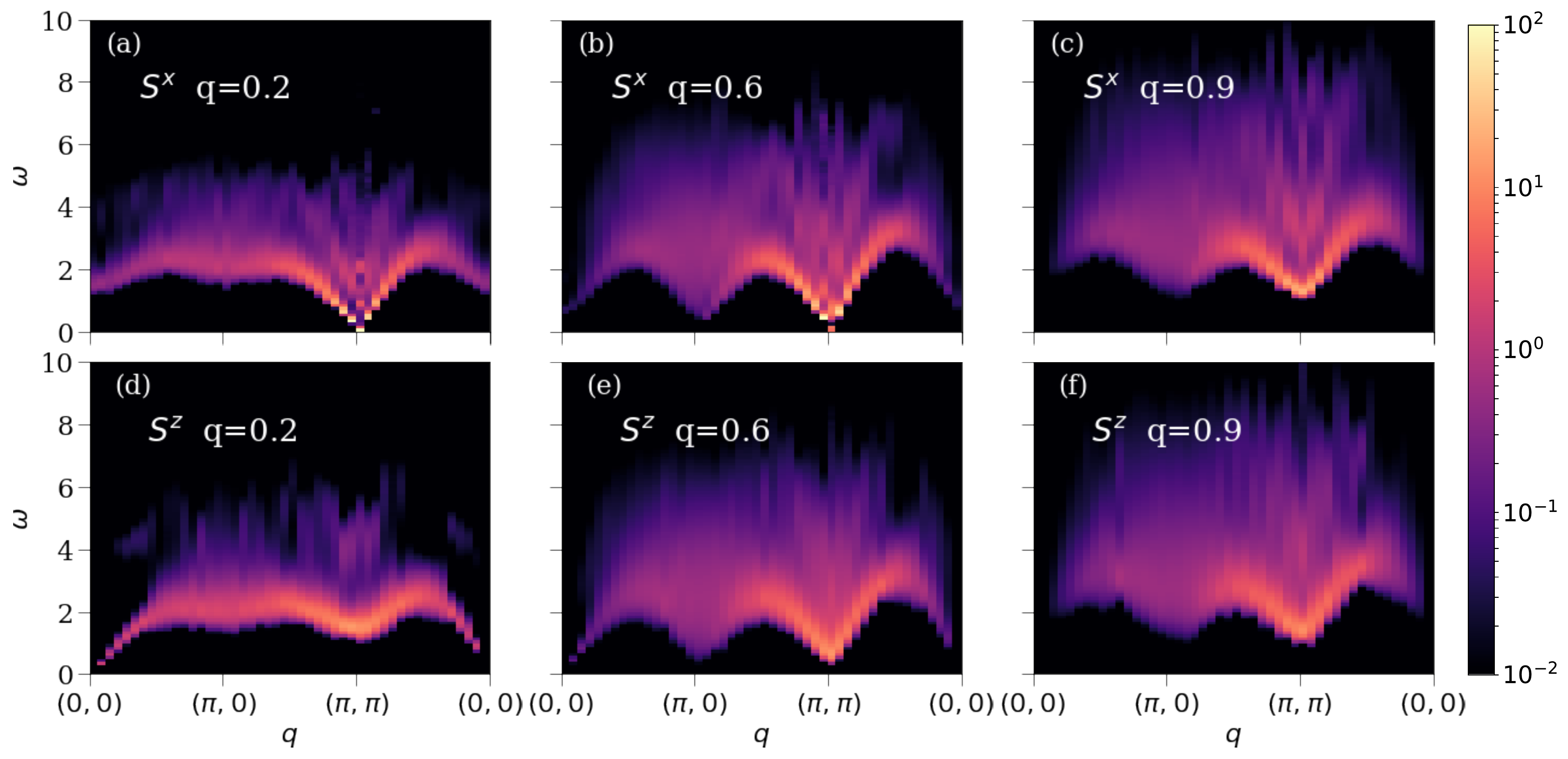}
\caption{Dynamic spin structure factors $S^{x}(\vect{q},\omega)$ (a-c) and
$S^{z}(\vect{q},\omega)$ (d-f) obtained from QMC-SAC calculations for the EPJQ 
model with $L=32$ and $\beta=64$. Here (a) and (d) are inside the AFXY phase, $q=0.2$, 
(b) and (e) are close to the DQCP, $q=0.6$, and (c) and (f) are inside the VBS phase, 
$q=0.9$.}
\label{fig:EPJQ32}
\end{figure*}

\begin{figure*}[htp!]
\includegraphics[width=\textwidth]{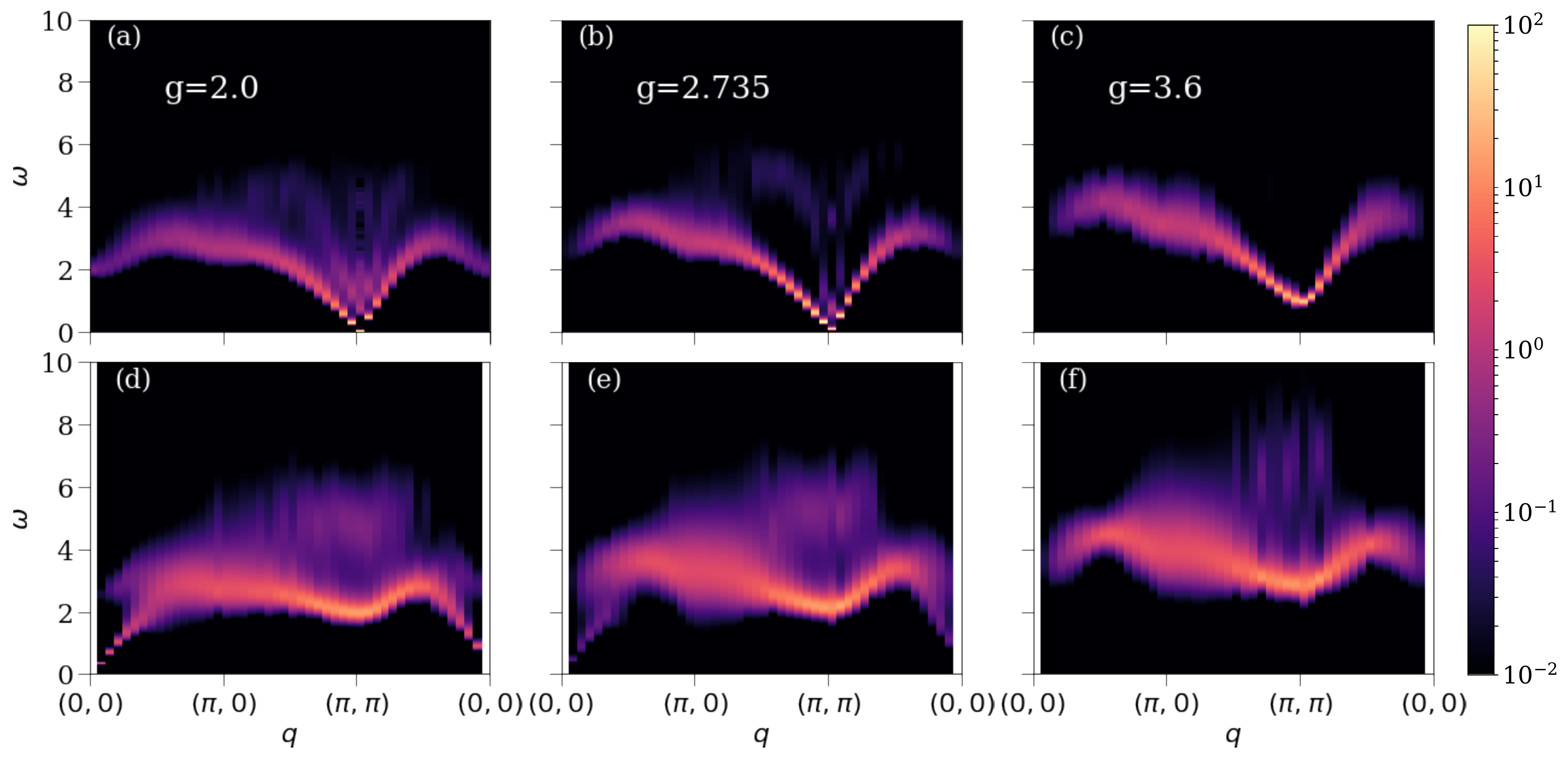}
\caption{Dynamic spin structure factors $S^{x}(\vect{q},\omega)$ (a-c) and
$S^{z}(\vect{q},\omega)$ (d-f) obtained from QMC-SAC calculations for the EPJ$_1$J$_2$
model with $L=32$ and $\beta=64$. Penels (a) and (d) are inside the AFXY phase, $g=2$, (b) and 
(e) are close to the 3DXY transition point, $g=2.735$, and (c) and (f) are inside the quantum
disordered phase, $g=3.6$.} 
\label{fig:EPJ1J232}
\end{figure*}

We have extracted $S(\vect{q},\omega)$ for the EPJQ and EPJ$_1$J$_2$ models in both the $S^x$ and $S^z$ channels. The imaginary time correlations 
in these channels were independently calculated in different simulations with the stochastic series expansion QMC method \cite{Sandvik1999}, 
implemented in the basis of the $S^x$ and $S^z$ spin components, respectively (i.e., the operators used in the correlators are always diagonal). 
For the EPJQ model in \eqnref{eq:EPJQ}, we set $J+Q=1$ and define the ratio $q=Q/(J+Q)$ to be the driving parameter, and for the EPJ$_1$J$_2$ 
model we define $g=J_2/J_1$. All results presented here are for $L\times L$ square lattices with $L=32$ and periodic boundary conditions, 
and the inverse temperature $\beta=1/T=2L$. While there are some remaining finite size effects on these lattices, by comparing with smaller 
lattices we have confirmed that the main features of the spectra are stable and should be close to the thermodynamic limit and $T=0$. 

For small $q$, the EPJQ model essentially reduces to an XXZ model, which has an AFXY ground state that breaks the $\text{U}(1)_\text{spin}$ 
symmetry spontaneously. When $q$ is large, the dimer interaction favors a VBS (columnar-dimerized) ground state, which spontaneously breaks the 
lattice $C_4$ rotation symmetry. In this work, we set the anisotropy parameter to $\Delta=1/2$, where we have found the signature of an easy-plane 
version of the DQCP separating the AFXY and the VBS phases at $q_c=0.6197(2)$, based on the finite-size analysis of the critical exponent in 
our previous work~\cite{YQQin2017}. The phase diagram of the EPJQ model \figref{fig:model}(a) is similar to those of the 
$\text{SU}(2)_{\text{spin}}$ $J$-$Q_2$ and $J$-$Q_3$ models \cite{Sandvik2002,Sandvik2007,Lou2009,Sen2010,Sandvik2010,Shao2016}, 
but the DQCP is in a different universality class due to the lowered symmetry. In the EPJ$_1$J$_2$ model, as $g$ increases there is a O(2) 
Wilson-Fisher transition from the AFXY phase to the trivial columnar dimer phase, as illustrated in \figref{fig:model}(b). The critical point is 
at $g_c=2.735(2)$ as determined in Appendix~\ref{sec:EPJ1J2QCP}. The O(3) vesion of this quantum phase transition has been investigated extensively 
with various statically dimerized Heisenberg Hamiltonians (see the review in Ref.~\onlinecite{Sandvik2010b}), including also a recent calculation 
of dynamic spectral functions \cite{Lohoefer2015}. The O(2) transition, however, has not been investigated in detail with 2D quantum spin 
models, as far as we are aware.

Turning now to the salient features of the spin spectra, in \figref{fig:EPJQ32} and \figref{fig:EPJ1J232} we show our results for the two models 
along the high symmetry path of wavevectors $(0,0)$-$(\pi,0)$-$(\pi,\pi)$-$(0,0)$. In both cases we present results both inside the two phases and at the critical point. 

In the AFXY phase, which is common to both the EPJQ and the EPJ$_1$J$_2$ models, we observe the gapless Goldstone mode at $(\pi,\pi)$ in the 
$S^x$ channel, as shown in \figref{fig:EPJQ32} (a) and \figref{fig:EPJ1J232} (a). In the $S^z$ channel, the $(\pi,\pi)$ fluctuations are gapped due to the easy-plane anisotropy. However, as seen in \figref{fig:EPJQ32} (b) and \figref{fig:EPJ1J232} (b), the modes around $(0,0)$ are 
still gapless, but with vanishing spectral weight as $\omega\to0$, as expected due to the conserved total $S^z$. These behaviors are consistent 
with theoretical expectations in \figref{fig:features} (a,b) based on the field theory of the XY model.

Now let us focus on the critical points of both models. \figref{fig:EPJQ32} (b,e) show the EPJQ spectra at $q=0.6$, close to the DQCP at 
$q_c=0.6197(2)$. \figref{fig:EPJ1J232} (b,e) show the EPJ$_1$J$_2$ spectra at $g=2.735$, close to the 3DXY transition at $g_c=2.735(2)$. By 
comparison, several exotic features of the DQCP spectra can be unambiguously identified. First, we observe broad and prominent continua in 
both $S^{x}(\vect{q},\omega)$ and $S^{z}(\vect{q},\omega)$ at the DQCP, which reflect the expected magnon fractionalization and emergence of deconfined, essentially 
independently propagating spinons. In contrast, at the 3DXY transition, the gapless magnon mode remains sharp in $S^{x}(\vect{q},\omega)$ around $(\pi,\pi)$ 
with very weak continuum due to the critical fluctuations. 

In the case of the DQCP, we find that the lower edges of both spectral function can be well accounted for by a remarkably simple single-spinon
dispersion relation, $\omega_1(\vect{q}) \propto [\sin^2(q_x)+\sin^2(q_y)]^{1/2}$, which matches the dispersion relation of a deconfined fermionic 
parton in the square lattice $\pi$-flux state \cite{hermele2005,senthilfisher,ranwen,Assaad2016}. This points us to the $N_f=2$ QCD or the $N_f=4$ QED 
theory \cite{SO5,youSMG2,2018arXiv180707574X} that were previously proposed to describe the DQCP. If indeed the broad spectral functions seen in \figref{fig:EPJQ32} (b,e)
are due to two independently propagated spinons, and contributions from four- or more spinon excitations can be neglected, the upper spectral bound is obtained
by maximizing $\omega_2(\vect{q})=\omega_1(\vect{q}_1)+\omega_1(\vect{q}_2)$ with $\vect{q}_1+\vect{q}_2=\vect{q}$. This indeed
appears to be in reasonable agreement with the observed distribution of the main spectral weight, though some weight, presumably arising from states
with more than two spinons, is also present at higher energies. As we will elaborate further in Sec.~\ref{sec:meanfield}, the overall shape
and weight distribution of the continuum can be nicely captured at the mean field level.

Note that the gapless continuum in $S^{z}(\vect{q},\omega)$ around $(\pi,\pi)$ is present at the DQCP but is absent at the 3DXY transition. The $S^z$ 
excitations are simply the low-energy fluctuations of the $n_z$ field. In the NCCP$^1$ description of the DQCP\cite{deconfine1,deconfine2}, 
$\scL[z,a]=|(\partial-\ii a - \ii A_s\frac{\sigma^z}{2})z|^2+\frac{\ii}{2\pi} A_v \wedge\dd a + \cdots$, the $n_z$ fluctuation corresponds to the two-spinon 
excitation since $n_z\sim z^\dagger \sigma^z z$. The criticality of the deconfined spinon $z$ therefore leads to the gapless $S^{z}(\vect{q},\omega)$ continuum around $(\pi,\pi)$.
However for the 3DXY transition, described by the XY 
order parameter $n_x+\ii n_y$ in the LGW paradigm, the $n_z$ fluctuation is not associated with any critical fluctuation of the order parameter 
and thus remains gapped at the critical point. Moreover, at the DQCP, as a consequence of the  $n_z$ criticality, the $S^x$ fluctuation also 
becomes gapless around $(0,0)$, because this corresponds to spin density $n_y\partial_t n_z-n_z\partial_t n_y$ which can decay into the gapless 
continuum of both $n_z$ and $n_y$. 

Furthermore gapless continua are also observed at momentum $(\pi,0)$ [and at $(0,\pi)$ as well by symmetry] in both $S^{x}(\vect{q},\omega)$ and $S^{z}(\vect{q},\omega)$ 
at the DQCP only. Gapless excitations at these points should be generally expected at a continuous quantum phase transition of the AFM into a columnar VBS 
state \cite{Spanu2006}, and the vanishing $(\pi,0)$ gap was already confirmed in the standard spin-rotational invariant [SU(2)] $J$-$Q$ model \cite{Suwa2016}.
It was also argued recently that the $(\pi,0)$ excitation anomaly (i.e., a lowered excitation energy and enhanced excitation continuum above the single-magnon
pole at this momentum relative to spin wave theory) in the Heisenberg model and, in materials realizing it, is a precursor to a DQCP \cite{Shao2016}.
Such exotic features are not observed at the 3DXY transition, however.

The ability to separate the distinct planar and out-of-plane excitations within the EPJQ model provides additional information and opportunities to 
test field-theoretical descriptions.
The $S^x$ excitation around $(\pi,0)$ corresponds to the fluctuation 
of the conserved current $n_x\partial_y v_y-v_y\partial_yn_x$ associated with the emergent O(4) symmetry (in the XY-VBS rotation channel), which 
is an unique feature of the easy-plane DQCP. The gapless point $(\pi,0)$ also follows naturally, because the XY-VBS current can decay into the 
$n_x$ continuum at $(\pi,\pi)$ and the $v_y$ continuum at $(0,\pi)$, such that the momenta add up to $(\pi,0)$. A similar interpretation applies 
to the $S^z$ channel as well. The only difference is that the spin-VBS current there is not conserved, but is nevertheless still critical. The $(\pi,0)$ continua exhibit a remarkable spatial anisotropy. On the edge of the continua, the spectral weight is always larger along $(\pi,0)$-$(\pi,\pi)$ line and smaller along $(\pi,0)$-$(0,0)$ line. This spatial anisotropy is a signature of current-current correlation, which originates from the non-trivial $\omega^2-q_x^2$ form factor on the numerator as given in \eqnref{eq:current1} and \eqnref{eq:current2}. The $(0,\pi)$ continua will also exhibit the spatial anisotropy but with the form factor rotated by $\pi/2$ to $\omega^2-q_y^2$. These ``shadow'' continua allow us to probe the critical VBS fluctuation in the spin excitation spectrum, which is another remarkable hallmark 
of the DQCP. 

As discussed in the Sec.~\ref{sec:introduction}, the spectral features uncovered here are relatively easy to probe in INS or RIXS experiments, 
hence paving way for observation of the seeming ephemeral DQCP in real materials.
These features are also robust even if the parameter is slightly off the critical point.
Our simulation itself serves as a ``numerical proof'' of this statement. As we measure the DQCP spectra at $q=0.6$ of the EPJQ model (not exactly at its critical point $q_c=0.6197(2)$), we still observe all the low-energy spectral features consistent with the field theory qualitatively. This demonstrates that the dynamical signatures do not require fine-tuning and should be easier to measure in experiment.
Whereas the previous studies of DQCP mainly focused on the 
critical scaling and exponents from the theoretical perspective, these quantities require more fine-tuning and are rather difficult to measure in experiments. Even if 
the DQCP turns out to be first-order (as expected if the anisotropy is strong) or becomes unstable against other intermediate phases at low 
temperature, its distinct spectral features over a large range of frequencies can still be robustly observed above the low energy scale at which 
the potentially other transitions of phases become manifest.

Finally, the spectra of the EPJQ model in the VBS phase is shown in \figref{fig:EPJQ32} (c,f). Their EPJ$_1$J$_2$ counterpart in the columnar 
singlet phase is shown in \figref{fig:EPJ1J232} (c,f). All spin excitations are gapped in both $S^{x}(\vect{q},\omega)$ and $S^{z}(\vect{q},\omega)$ for both models. For the 
EPJQ model, the spectra in the VBS phase still maintain broad continua above the gap, in contrast to the much sharper spectra of gapped magnons 
in the EPJ$_1$J$_2$ columnar phase. This might be related to the two-length-scale phenomena, which is inherent to the DQCP, persisting in the 
VBS phase of the standard JQ model~\cite{Shao2016}, namely, the domain wall size of the VBS order may still remain large while the spin correlation 
length is small. The domain wall size of the VBS order is directly related to the confinement length scale of the spinons \cite{deconfine2}. This implies 
that although the spin correlation length is finite, the confinement length scale of the spinon can still be large, which leads to the large continuum 
above the spin gap in the spin excitation spectrum. 

\section{Parton Mean Field Theory for the DQCP Spectra}
\label{sec:meanfield}
\begin{figure}[htp!]
\begin{center}
\includegraphics[width=\columnwidth]{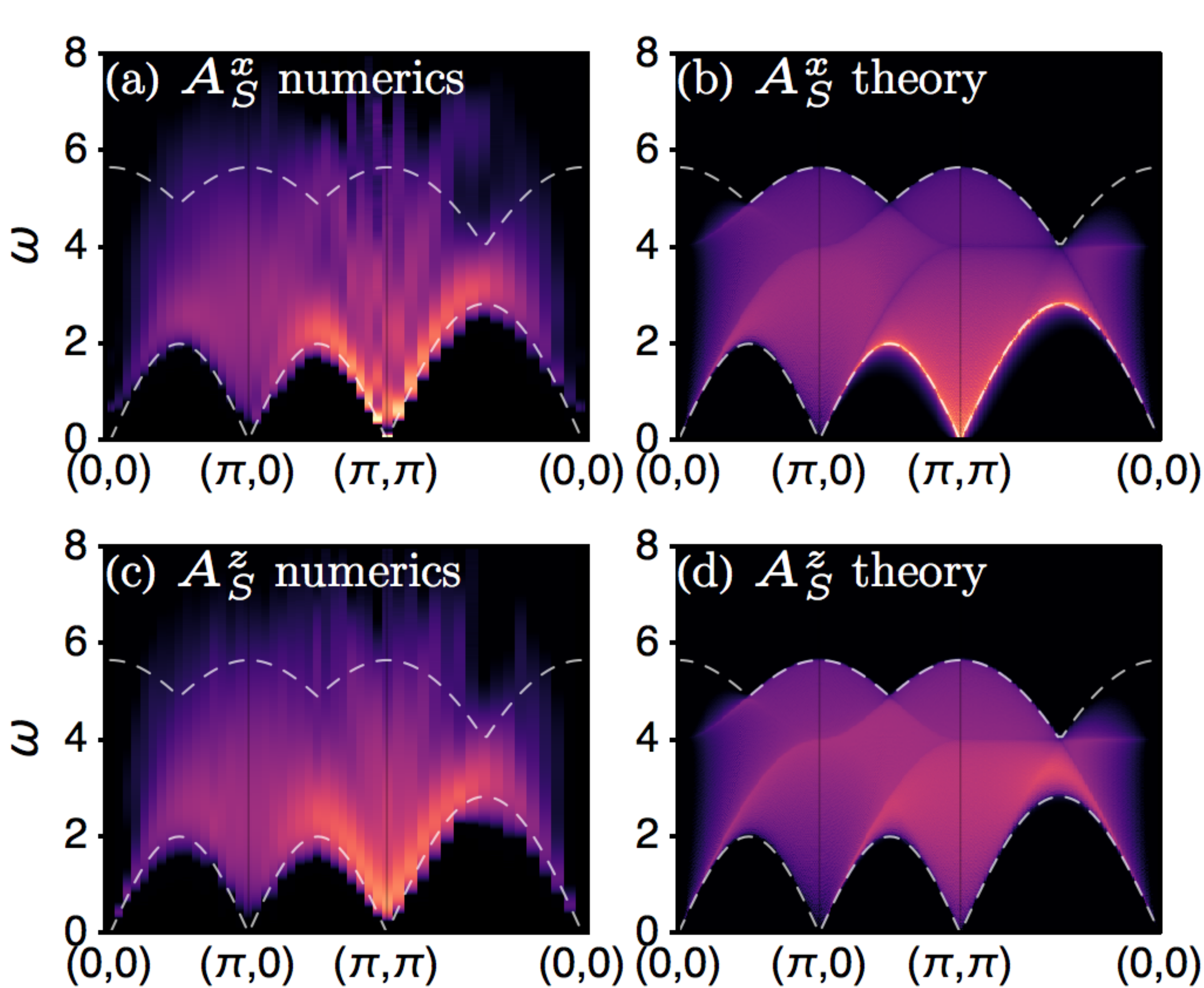}
\caption{Comparison of the DQCP dynamic spin structure factors between numerics [(a) $S^{x}$ channel and (c) $S^{z}$ channel] and theory [(b) $S^{x}$ channel and (d) $S^{z}$ channel]. The color map is the same as that in \figref{fig:EPJQ32}. The dashed 
curves trace out the upper and low edges of the two-parton continuum, assuming free fermionic partons with the $\pi$-flux state dispersion 
$\epsilon_{\vect{k}}$ in \eqnref{eq:dispersion}. The lower edge simply follows $\epsilon_{\vect{k}}$ and the upper edge is given by the maximal 
two-parton excitation energy $E_{\vect{q}}=\max_{\vect{k}\in\text{BZ}}|\epsilon_{\vect{k}}+\epsilon_{\vect{q}-\vect{k}}|$. The suppressed spectral 
weight near $(0,0)$ can be captured by matrix element effects.} 
\label{fig:compare}
\end{center}
\end{figure}

In this section, we provide theoretical account for the overall shape of the dynamic spin structure factors $S^{x}(\vect{q},\omega)$ and $S^{z}(\vect{q},\omega)$ 
observed at the DQCP. The easy-plane DQCP admits several candidate field theory descriptions, including the easy-plane NCCP$^1$ 
theory \cite{ashvinlesik,deconfine1,deconfine2}, the $N_f=2$ non-compact QED$_3$ 
theory \cite{senthilfisher,groverashvin,xudual,karchtong,seiberg2,mengxu,SO5} and the $N_f=2$ QCD$_3$ theory \cite{ranwen,SO5} 
(or its Higgs descendent $N_f=4$ compact QED$_3$ \cite{hermele2005,senthilfisher,SO5,youSMG2,2018arXiv180707574X}) with additional anisotropy in the SO(5) 
symmetric tensor representation. Although all theories are believed to provide equivalent descriptions of the low-energy physics under 
proposed duality relations \cite{SO5}, some of them are more convenient to handle by mean field treatment than others. Among these theories, 
we found that the $N_f=2$ QCD (or $N_f=4$ QED) theory gives the best account for the overall spectral features at the mean field level. 
Because in these theories, both the AFM and VBS order parameters are treated on equal footing as fermionic parton bilinears, it is already 
possible to approximately capture both spin and dimer fluctuations at the free fermion level (ignoring gauge fluctuations and local 
interactions). \figref{fig:compare} shows the comparison of the dynamics spin structure factors between numerics and theory, based on 
the parton mean field theory. The overall features match quite nicely. However, if similar mean field treatment were applied to other dual field theories such as the NCCP$^1$ or the $N_f=2$ non-compact QED$_3$ theories, some low-energy continua that involve gauge monopole excitations will be missing, as the gauge fluctuation can not be captured at the mean field level.

Let us start with the parton construction on the square lattice \cite{wenbook}, where the spin operator $\vect{S}_i$ is fractionalized 
into fermionic partons $f_{i}=(f_{i\uparrow},f_{i\downarrow})^\intercal$ at each site $i$ as
\begin{equation}\label{eq:S=ff}
\vect{S}_i=\frac{1}{2}f_i^\dagger \vect{\sigma} f_i.
\end{equation}
An SU(2) gauge structure emerges in association with the above fractionalization scheme, but at the mean field treatment we will ignore 
the SU(2) gauge fluctuation completely and place the fermionic parton in the square-lattice $\pi$-flux state \cite{wenbook,hermele2005,ranwen}.
Thus, we use the following mean field Hamiltonian
\begin{equation}\label{eq:HMF}
H_\text{MF}=\sum_{i}\ii(f_{i+\hat{x}}^\dagger f_i+(-)^x f_{i+\hat{y}}^\dagger f_i)+\text{H.c.},
\end{equation}
such that each plaquette hosts a $\pi$-flux for the fermionic parton. Four Dirac fermions are obtained at low energy. The fermionic parton 
dispersion is simply given by
\begin{equation}
\label{eq:dispersion}
\epsilon_{\vect{k}}=2(\sin^2(k_x)+\sin^2(k_y))^{1/2}.
\end{equation}
It is interesting to find that the lower edge of the DQCP spectra follows this simple dispersion relation quite nicely without any adjustable
parameters beyond an overall velocity, as shown in \figref{fig:compare} (a,c), which justifies the $\pi$-flux state as our starting point. The 
upper edge of the two-parton continuum can also be obtained from $\epsilon_\vect{k}$ by adding up single-parton energies. This gives a rough 
estimate for the energy range of the parton continuum, which is also consistent with the numerical observation in \figref{fig:compare} (a,c). 

\begin{figure}[htbp]
\begin{center}
\includegraphics[width=0.9\columnwidth]{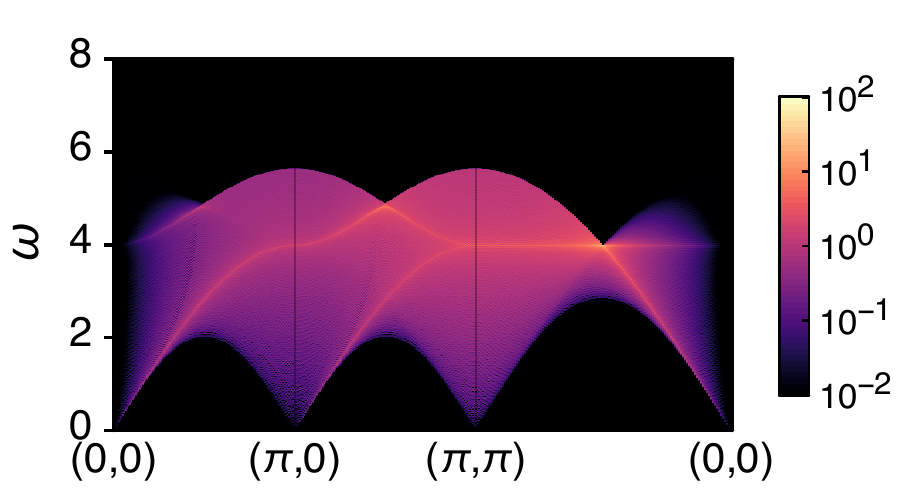}
\caption{The (bare) dynamic spin structure factor $S_0(\vect{q},\omega)$ of the free fermion $\pi$-flux state.}
\label{fig:bare}
\end{center}
\end{figure}

Given \eqnref{eq:S=ff} and \eqnref{eq:HMF}, it is straightforward to calculate the spin-spin correlation function
\begin{equation}
G_0^{a}(\vect{r}_i-\vect{r}_j,t)=\langle\text{MF}|e^{\ii H_\text{MF} t}S_i^a e^{-\ii H_\text{MF}t} S_j^b|\text{MF}\rangle
\end{equation}
on the free fermion ground state $|\text{MF}\rangle$ of the mean field Hamiltonian $H_\text{MF}$. Then we can obtain the dynamic spin 
susceptibility 
\begin{equation}
\chi_0^{a}(\vect{q},\omega)=\int\dd t\sum_{i}G_0^{a}(\vect{r}_i,t)e^{\ii\omega t-\ii\vect{q}\cdot\vect{r}_i},
\end{equation}
from which we obtain the dynamic spin structure factor 
\begin{equation}
S_{0}^{a}(\vect{q},\omega)=\Im \chi_0^{a}(\vect{q},\omega+\ii 0_+),
\end{equation}
graphed in \figref{fig:bare}. This spectral function was also calculated in Ref.\,\onlinecite{Assaad2016} previously. One can see that
$S_0$ already captures the gapless continua at momenta $(0,0)$, $(\pi,0)$, $(0,\pi)$, and $(\pi,\pi)$ in all spin channels. Because the mean 
field Hamiltonian $H_\text{MF}$ is symmetric under SU(2)$_\text{spin}$, there is no difference between $S_0^{x}(\vect{q},\omega)$ and $S_0^{z}(\vect{q},\omega)$. 
The easy-plane anisotropy only enters the parton theory starting from four-fermion interactions, since it is expressed in the SO(5) symmetric tensor
 representation that can not be written down at the quadratic level. Therefore, the anisotropy is not manifest in the mean field approximation, 
where the interaction effects are ignored. This observation provides a natural explanation for the strikingly similar spectra of $S^{x}(\vect{q},\omega)$ and $S^{z}(\vect{q},\omega)$ seen in the numerical results in Sec.~\ref{sec:spectra} at the DQCP, despite of the presence of a rather large anisotropy $\Delta=1/2$ 
in the EPJQ model.

The gauge fluctuations are expected to further renormalize 
the spectrum and enhance the critical fluctuations around $(\pi,\pi)$, which are not taken into account in the simple mean field theory presented in \figref{fig:bare}. While including the gauge interactions in the calculation is highly non-trivial and  beyond the scope of this work, we next discuss a phenomenological model that captures the spectral weight enhancement, and leave the more extensive calculation to future work. Let us consider modeling the interaction effect phenomenologically by a random phase approximation (RPA) 
correction,
\begin{equation}
\chi^{a}(\vect{q},\omega)=\frac{\chi_0^{a}(\vect{q},\omega)}{1+J_a \chi_0^{a}(\vect{q},\omega)},
\end{equation}
where $a=x,y,z$. The coupling $J_a$ parameterize the strength of the spin-spin interaction in the $S^a$ channel. We can introduce the easy-plane anisotropy simply by considering $J_x=J_y>J_z$. We found that the $(\pi,\pi)$ fluctuation is indeed enhanced by the interaction $J_a$. The resulting RPA corrected spectral functions are already shown in \figref{fig:compare} (b,d), with $J_x$ tuned to the magnetic ordering critical point and $J_z=J_x/2$ \footnote{Although such a Grose-Neveu critical point is different from the DQCP, we only use it to provide a rough estimate of the spectral features close to a magnetic ordered phase. We do not claim that the criticality of DQCP can be correctly understood by our mean field + RPA approach.}. Compared to \figref{fig:bare}, the spin spectra in \figref{fig:compare} (b,d) are much improved by the interaction effect. Our phenomenological study combined with the QMC-SAC result demonstrates that the $\pi$-flux state fermionic parton with interaction accounts well for the overall features of the DQCP spectra in both $S^x$ and $S^z$ channels, which is consistent with the expectations from the $N_f=2$ QCD or $N_f=4$ QED theories. An interesting open problem is a systematic route to incorporating the effects of gauge fluctuations in calculating the spin excitation  spectrum.

\section{Discussion}
\label{sec:discussion}

In this work, we have demonstrated dynamical signatures of fractionalization at the DQCP in a planar, U(1), quantum magnet by computing both the in-plane and 
out-of-plane dynamic spin structure factors at low temperature. By contrasting with analogous results for a conventional LGW critical point, we explicitly 
observe how fractionalization of the critical magnon into two spinons is manifested by a large continuum, in sharp contrast to a much less prominent continuum 
due to conventional critical quantum fluctuations at the ordinary 3DXY transition. We also discovered several low-energy spectral features that are unique to 
the DQCP, notably the $(\pi,\pi)$ continuum in the $S^z$ channel and the $(\pi,0)$ continua in both channels. These features are missing in the 3DXY transition. 
They will provide us with ``smoking gun'' evidence to guide experimental searches for DQCPs. In particular, the calculations we have presented here can 
be compared with neutron scattering experiments on quantum magnets with weak antiferromagnetic (or similar) order or quantum paramagnets with small gaps. 

An important aspect of experimental detection of DQCP physics is that we have demonstrated the salient dynamical signatures of confinement, in the form of 
broad continua of spin excitations, even quite far away from the DQCP, inside the gapped phase. This reflects an expected long length scale associated with 
deconfinement inside the VBS state, which may make it possible to observe essentially deconfined spinons even quite far from DQCP in quantum paramagnets with 
the right kind of fluctuations. Some signatures of precursors to a DQCP have already been argued on the gapless side, in materials closely realizing the 2D 
Heisenberg antiferromagnet \cite{Shao2017b}, and in systems with weakened order these signatures should become stronger; again under the condition that the
fluctuations are those associated with the DQCP, instead of bringing the system closer to some other exotic phase \cite{Chatterjee17}. One promising candidate
for DQCP physics is the magnetically quasi-2D Shastry-Sutherland material {S\lowercase{r}C\lowercase{u}$_2$(BO$_3$)$_2$}, where a plaquette singlet state similar 
to a VBS was recently detected by neutron scattering at moderately high pressure \cite{Zayed17}, before the system enters an antiferromagnetic phase at
higher pressure. Although in this case the non-magnetic state is two-fold degenerate, instead of the four-fold degenerate VBS states we have considered 
here, a DQCP with emergent O(4) symmetry may still be realizable close to the parameter regime of the material \cite{Zhao18}.

Our work also calls for further studies on the theory side. While we have found remarkable agreement between the numerical spectra and field theories of
deconfined spinons at the mean field level, there are also distinctive features, e.g., the way the spectral weight at the lower edge evolves as we move through 
the BZ, that will require more sophisticated treatments of the interactions and gauge fluctuations. Our results should provide a concrete impetus for these 
demanding calculations. Beyond quantum magnets, in a series of recent works~\cite{Karch2017,potterdual,SO5}, it was shown that the easy-plane DQCP is dual to the 
$N=2$ quantum electrodynamics with fermionic matter fields, and it describes the bosonic topological transition (BTT) between a bosonic symmetry protected 
topological state and a trivial insulator state~\cite{groverashvin,lulee}. In addition, it also exhibits a self-duality~\cite{xudual,Karch2017,mrossduality,seiberg2}. 
Recently this bosonic topological transition has also been realized in a lattice model and the static properties of the phase transition has been simulated 
via determinantal QMC~\cite{kevinQSH,mengQSH2,Wu2016,yoshida2016}. It would be very interesting to measure the dynamic structure factor at the BTT, and 
compare the results with our current study.

The DQCP also has a natural large-$N$ generalization \cite{deconfine1,deconfine2}, i.e., instead of two flavors of bosonic matter fields, there are $N$-flavors 
of matter fields. The field theory is called the noncompact CP$^{N-1}$ model. This model is understood very well in the large-$N$ limit, and for finite and large 
$N$, a systematic $1/N$ expansion can be performed to understand the details of the NCCP$^{N-1}$. Remarkable agreements have already been found between critical
exponents calculated in the $1/N$ expansion \cite{Dyer2015,Dyer2016} and results of finite-size scaling of QMC data \cite{Kaul12,Block2013,Jonathan2017}. In the future,
similar dynamic structure factor calculations for the large-$N$ versions of the DQCP and comparison with QMC-SAC calculation would be another very interesting 
research direction. 

\acknowledgments 
The authors thank Fakher Assaad, Yin-Chen He, Max Metlitski, Subir Sachdev, Hui Shao, T. Senthil, Chong Wang and Stefan Wessel for helpful discussions.  N.S.M., G.Y.S. and Z.Y.M. are supported by the Ministry of Science and Technology of China under Grant No. 2016YFA0300502, the Strategic Priority Research Program  of the Chinese Academy of Sciences (XDB28000000) and the National Science Foundation of China under Grants No. 11421092, No. 11574359 and No. 11674370. NSM would like to thank Boston University for support under its Condensed Matter Theory Visitors Program. AV and YZY are supported by a Simons Investigator Grant. AWS is supported by the NSF under Grant No. DMR-1710170 and by a Simons Investigator Grant. CX is supported 
by the David and Lucile Packard Foundation and NSF Grant No.~DMR-1151208. We thank the following institutions for allocation of CPU time: the 
Center for Quantum Simulation Sciences at the Institute of Physics, Chinese Academy of Sciences, and the Tianhe-1A platform at the National 
Supercomputer Center in Tianjin. 

\bibliographystyle{apsrev4-1}
\bibliography{spectra_epjq}

\appendix
\section{Quantum critical point of the EPJ$_1$J$_2$ model}
\label{sec:EPJ1J2QCP}
\begin{figure}[htp!]
\includegraphics[width=\columnwidth]{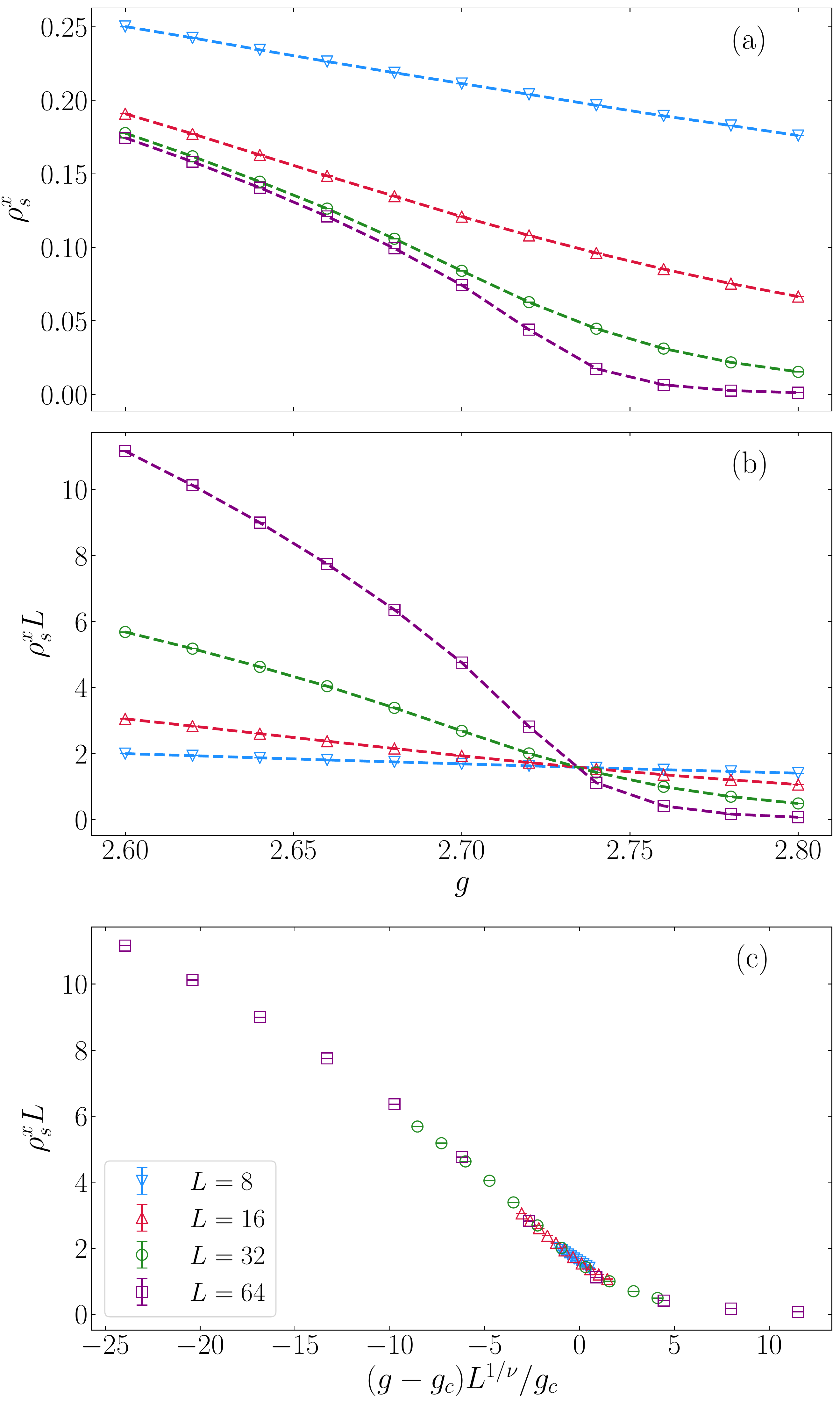}
\caption{Finite-size scaling of the spin stiffness $\rho^{x}_s$
of the EPJ$_1$J$_2$ model. (a) Raw data for system sizes
$L=8,16,32$ and $64$ as a function of $g={J_2}/{J_1}$. (b) The
same results rescaled by $L$, so that the crossing points of
the curves for different $L$ should approach the critical point,
determined here as $g_c=2.735\pm0.002$. (c) The horizontal axis
has been further rescaled as $(g-g_c)L^{1/\nu}$ with the 3D O(2) 
exponent $\nu$, leading to good data collapse in support of the
assumed universality class.}
\label{fig:rhos}
\end{figure}

In this section, we discuss how the  critical point $g_c={J_2}/{J_1}=2.735(2)$ of the EPJ$_1$J$_2$ model is determined from finite-size scaling of
QMC results. The physical observable used here is the spin stiffness
\begin{equation}
\rho_{s}\equiv \frac{1}{N}\frac{\partial^{2}F(\phi)}{\partial\phi^{2}},
\end{equation}
where $F$ is the free energy and $\phi$ is the twisting angle between
spins in two different columns.

$\rho_{s}$ is easy to measure in SSE QMC simulations \cite{Sandvik2010b} and it has well defined finite size scaling form
\begin{equation}
\rho_{s}(g,L)=L^{-z}f((g-g_c)L^{1/\nu},{\beta}{L^{-z}}),
\label{eq:rhofss}
\end{equation}
with the dynamic exponent $z=1$ and the correlation length exponent $\nu=0.672$, and $f(x,y)$ is the scaling function. The quantum phase transition from AFXY phase to the columnar phase in the EPJ$_1$J$_2$ model belongs to $(2+1)$D O(2) universality class ~\cite{Hasenbusch1999,Meng2008}. In the simulation we fix $\beta=2L$ such that the ${\beta}/{L^{-z}}$ is a constant and we effectively have a single-parameter scaling function. Then the quantity $\rho_{s}L$ is dimensionless and 
does not change with $L$ at critical point. 

In the QMC simulation, $\rho_s$ is evaluated using winding number fluctuations as \cite{Sandvik2010b}
\begin{equation}
\rho_{s}= \frac{1}{N\beta}\langle(N^{+}-N^{-})^{2}\rangle,
\end{equation}
where $N^{+}$ and $N^{-}$ are the number of all operators that transport spin in positive and negative directions along the
lattice direction, respectively, during propagation of the spin state in imaginary time. Since the EPJ$_1$J$_2$ model is spatially anisotropic, 
one can calculate $\rho_s$ along the $x$ or $y$ directions of the lattice, and both of them satisfy the same finite size scaling form \eqnref{eq:rhofss} with 
different scaling functions. We label the two quantities as $\rho^{x}_{s}$ and $\rho^{y}_{s}$, and for the sake of simplicity, only show $\rho^{x}_{s}$ here.

\figref{fig:rhos} (a) depicts the raw $\rho^{x}_{s}$ data for four different system sizes, $L=8,16,32,64$. In \figref{fig:rhos}(b), we plot the scaled 
quantity $L\rho^{x}_{s}$ against the control parameter. The crossing point of the curves should drift toward the critical point $g_c$, and from our 
data we obtain $g_c=2.735(2)$. Finally, in \figref{fig:rhos}(c), we further rescale the $x$-axis as $(g-g_c)L^{1/\nu}$, with $g_c=2.735$ and $\nu=0.672$,
thus obtaining the scaling function in \eqnref{eq:rhofss} as the common curve onto which the data for different system sizes collapse. The good data
collapse without any adjustable parameters supports the expected $(2+1)$D O(2) universality class.

\section{Spectra of EPJ$_1$J$_2$ model}
\label{sec:EPJ1J2completespectra}

\begin{figure*}[htp!]
\includegraphics[width=\textwidth]{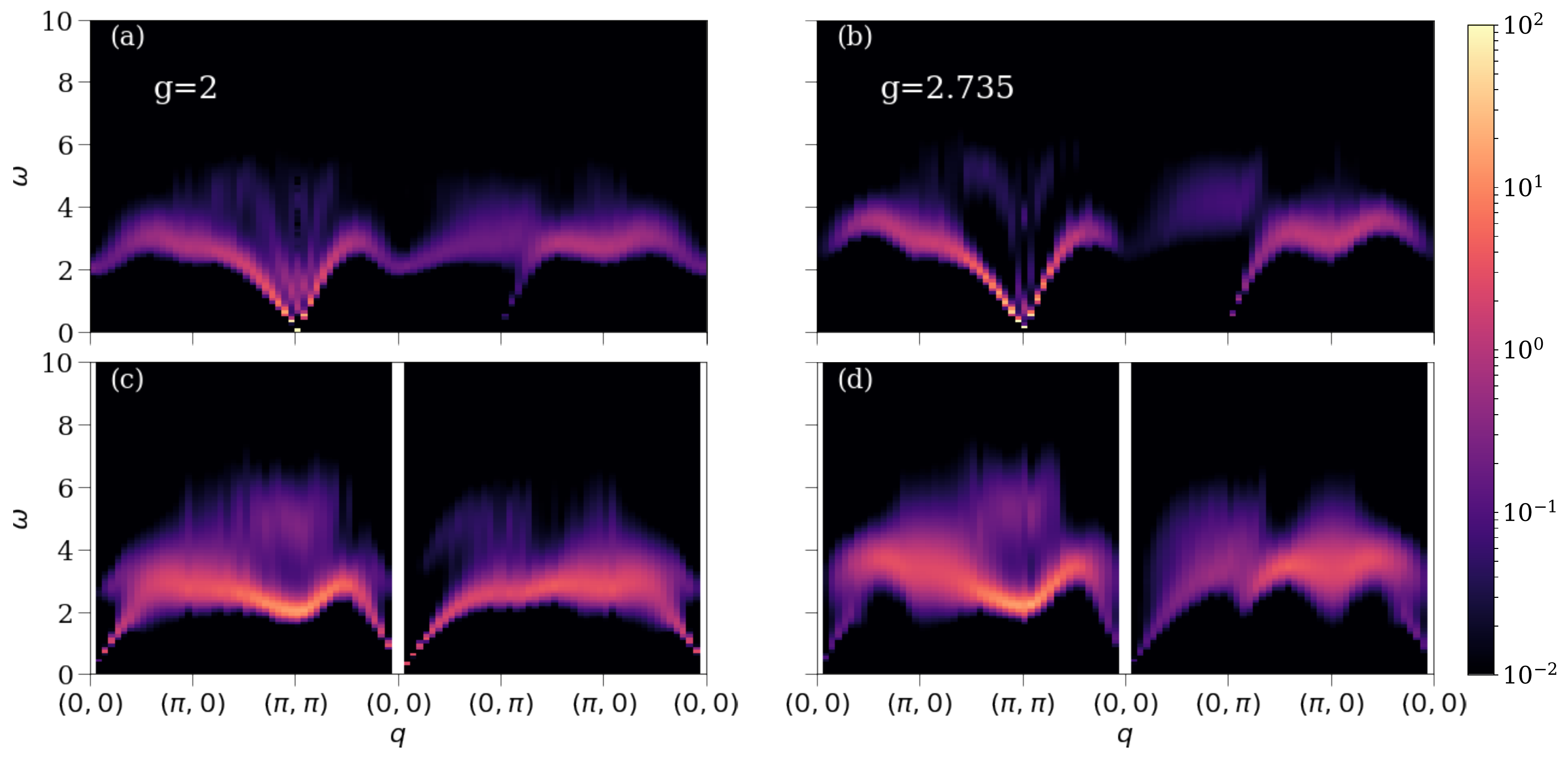}
\caption{Spin structure factors $S^{x}(\vect{q},\omega)$ and $S^{z}(\vect{q},\omega)$ obtained from QMC-SAC for the EPJ$_1$J$_2$ model with 
$L=32$ and $\beta=2L$. Panels (a) and (c) show data inside the AFXY phase with $g=2$, while (b) and (d) are close to the 3DXY transition point 
with $g=2.735$. Results are shown along the path $(0,0)-(\pi,0)-(\pi,\pi)-(0,0)-(0,\pi)-(\pi,0)-(0,0)$ through the BZ.}
\label{fig:EPJ1J2_32_complete}
\end{figure*}

In this section, we provide more detailed information of the EPJ$_1$J$_2$ spectra inside the AFXY phase and close to the 3DXY
transition point. \figref{fig:EPJ1J2_32_complete} shows a scan through the square-lattice Brilliune zone (BZ) with more $\vect{q}$ points, 
along the path $(0,0)-(\pi,0)-(\pi,\pi)-(0,0)-(0,\pi)-(\pi,0)-(0,0)$. In \figref{fig:EPJ1J2_32_complete}(a), $S^{x}(\vect{q},\omega)$ is 
shown at $g=2$. Here the left part is identical to \figref{fig:EPJ1J232}(a), where the Goldstone mode at $(\pi,\pi)$ is seen, and the spectra 
at $(\pi,0)$ is gapped. The right part is slightly different, with the spectra at $(0,\pi)$ also gapped, but, due to the folding of the BZ coming 
from the doubling of the real space unit cell  along the $x$ direction of the square lattice, a ``shadow'' band originating from the 
gapless dispersion from $(\pi,\pi) - (\pi/2,\pi/2)$ of the AFXY phase, presents itself from $(0,\pi) - (\pi/2,\pi/2)$. \figref{fig:EPJ1J2_32_complete}
(c) shows $S^{z}(\vect{q},\omega)$ along the the same path. Since the $S^{z}$ excitations are gapped in the AFXY phase, except for the $(0,0)$ point, 
there is no obvious sign of the band folding close to $(0,\pi)$.

\figref{fig:EPJ1J2_32_complete}(b),(d) show $S^{x}(\vect{q},\omega)$ and $S^{z}(\vect{q},\omega)$ close to the 3DXY critical point 
at $g=2.735$. The Goldstone mode at $(\pi,\pi)$ still presents itself, as well as the band folding close to $(0,\pi)$ in the $S^{x}$ channel. 
In the $S^{z}$ channel, on the other hand, all the spectra are gapped except $(0,0)$, and above the gap, some signature of the band-folding 
can also be seen close to $(0,\pi)$.

\end{document}